\documentclass[prl,twocolumn,showpacs,preprintnumbers,amsmath,amssymb]{revtex4}

\usepackage{dcolumn}
\usepackage{bm}
\usepackage{graphicx}

\begin{document}

\title{Optimized Interactions for Targeted Self-Assembly: Application to Honeycomb Lattice}

\author{Mikael Rechtsman$^1$}
\author{Frank Stillinger$^2$}
\author{Salvatore Torquato$^{2,3}$}
\affiliation{$^1$Department of Physics, Princeton University, Princeton, NJ 08544}
\affiliation{$^2$Department of Chemistry, Princeton University, Princeton, NJ 08544}
\affiliation{$^3$Program in Applied and Computational Mathematics and PRISM, Princeton, NJ 08544}

\date{\today}

\begin{abstract}
We devise an inverse statistical-mechanical methodology to find optimized interaction
potentials that lead spontaneously to a target many-particle
configuration. Target structures can possess varying degrees of
disorder, thus extending the traditional idea of self-assembly to
incorporate  both amorphous and crystalline structures as well
as quasicrystals. For illustration purposes, our computational technique is applied
to yield an optimized isotropic (non-directional) pair
potential  that spontaneously yields the three-coordinated honeycomb lattice as
the ground state structure in two dimensions.
This target choice is motivated by its three-dimensional
analog, the diamond lattice, which is known to possess
desirable photonic bandgap properties.

\end{abstract}


\maketitle

``Self-assembly" typically describes processes in which entities
(atoms, molecules, aggregates of molecules, etc.)  spontaneously
arrange themselves into a larger ordered and functioning
structure. Biology offers wonderful examples, including the
spontaneous formation of the DNA double helix from two complementary
oligonucleotide chains, the formation of lipid bilayers to produce
membranes, and the folding of proteins into a biologically active
state. Molecular self-assembly is a potentially
powerful method to fabricate atomically precise materials and devices.
For example, Whitesides \cite{whitesides1} has shown intricate
two-dimensional patterns can emerge in self-assembly of organic
molecules on an inorganic surface.  Jenekhe and Chen \cite{Jen99} have
devised `smart plastics' that assemble into photonic crystals.
Manoharan et. al. have self-assembled unique, small clusters of microspheres
\cite{Pine03}.  These examples provide glimpses into the materials
science of the future, i.e., devising building blocks with specific
interactions that can self-organize on a set of larger length scales.
Theoretical work has mainly focused on finding the structure and
macroscopic properties of many-body systems given the interactions --
what we refer to as the ``forward" problem of statistical
mechanics. The forward problem has been extensively studied in the
context of the freezing transition both analytically
\cite{ramakrishnan} and numerically \cite{Hansen}, and more recently
by Kamien \cite{kamien2}, who uses geometric arguments to obtain
crystal entropy.

The purpose of this Letter is to introduce an {\it inverse
statistical-mechanical} methodology to find optimized interaction
potentials that lead spontaneously to a target many-particle
configuration.  The so-called `reverse' Monte Carlo method
\cite{Laaksonen,Mueller-Plathe} has been used to obtain interactions in liquids given
the pair correlation function, which only has partial configurational information.  
Our inverse methodology distinguishes itself in that we
apply it to self-assembly of a given N-particle configuration, which
may be be crystalline, quasicrystalline, or amorphous.  We envision
target structures possessing varying degrees of disorder, which
enables us to extend the traditional idea of self-assembly.

The idea of {\it tailoring} potentials to generate targeted structures
is motivated by the rich array of fundamental issues and questions
offered by this fascinating inverse statistical-mechanical problem as
well as our recent ability to identify the structures that have
optimal or desirable bulk properties. The latter includes novel
crystal structures for photonic band-gap applications \cite{Ho90},
materials with negative or vanishing thermal expansion coefficients
\cite{Si96}, materials with negative Poisson ratios \cite{Xu99},
materials with optimal transport and mechanical properties
\cite{Hy01}, and mesoporous solids for applications in catalysis,
separations, sensors and electronics \cite{Fer99}.

Colloids are the ideal system to test our optimized potentials, since
both repulsive and attractive interactions can be manipulated (e.g.,
via depletion forces, dipole-dipole interactions, electrostatic
interactions, etc. \cite{Russel}) and therefore offer a panoply of
possible potentials that far extends the range offered by molecular
systems.

Because there is a vast class of many-body potentials, we will focus
on isotropic pairwise additive interactions for simplicity here. There
are many open questions even for this simple class of potentials. For
example, it is not known what are the limitations of isotropic
pairwise additivity for producing target structures. We know that such
interactions cannot produce thermodynamically stable chiral structures
with a specified handedness; equal amounts of left-handed and
right-handed structures would result.  When is anisotropy in the
potential required? An answer based on intuition from molecular
systems would fail here. For instance, the diamond lattice is thought
to require directional interactions because such structures found in
Nature result from covalent bonding.  In fact, it is not known whether
a diamond lattice could be created from an isotropic pair
potential. This structure has a special status in photonics research
because a diamond lattice of
dielectric spheres exhibits a photonic band gap across the Brillouin
zone \cite{Ho90}.

The two-dimensional analog of this open three-dimensional crystal is
the three-coordinated honeycomb lattice. Accordingly, our general
optimization procedure (described below) will be illustrated by
applying it to produce an optimized circularly symmetric pair
potential $V(r)$ that spontaneously yields the honeycomb lattice as
the ground state (zero-temperature) structure in a positive density
range. In contrast to previous approaches that have claimed to produce
open lattice structures, our procedure incorporates the phonon
spectra, which is a crucial ingredient.  Because the honeycomb is an
open lattice that is a subset of the triangular lattice, it is
inherently challenging to assemble using isotropic potentials. Indeed,
such a potential has never been found before.

The  potential energy for a system of $N$ classically interacting 
particles at positions $\bf r^N \equiv {\bf r}_1, {\bf r}_2, \ldots, {\bf r}_N$ in 
the absence of an external field is given by
\begin{equation}
\Phi({\bf r^N}) = \sum_{i<j}V_2({\bf r}_i,{\bf r}_j) + \sum_{i<j<k}V_3({\bf r}_i,{\bf r}_j,{\bf r}_k) + ...
\end{equation}
where the $V_n$'s are $n$-body potentials.  
In this study, we consider only isotropic pair potentials and therefore
\begin{equation}
\Phi({\bf r^N}) = \sum_{i<j}V(|{\bf r}_i - {\bf r}_j|).
\end{equation}


A central feature of our inverse approach is a
computational algorithm that searches for and optimizes a functional
form for $V(r)$ that leads to self-assembly of a given target
structure.  To find an optimized $V(r)$ for a given target structure, we make an
initial guess for that function. We require that this initial
potential have real frequencies for each of its normal modes
(for a lattice, this means real phonon frequencies for all wavevectors
in the Brillouin zone).  Thus, the structure is mechanically stable at
zero temperature.  We then parameterize the potential, establishing a
family of functions $V(r; \{a_0...a_n\})$ of which our initial guess
is a member.  The parameterization must be chosen so that an overall
rescaling is not possible.  For each $a_i$, we choose a range of values
that it can take, namely $[a_i^{min}, a_i^{max}]$.  We then optimize
this family of functions for self-assembly. Specifically, the program runs a molecular
dynamics simulation (MD) at volume (or area) per particle $\alpha$,
initially in the target structure configuration and interacting via
the initial guess potential.  The initial root mean square speed in
the MD corresponds to about 90-95\% of the melting temperature, where
the system is in a non-harmonic regime and liquid nucleation and/or
structural phase transition is beginning to set in.  The root mean
square deviation from the target structure, defined as
\begin{equation}
L = \sqrt{\frac{1}{N}\sum_i ({\bf r}_i - {\bf r}_i^{(0)})^2 - \left( \frac{1}{N}\sum_i ({\bf r}_i-{\bf r}_i^{(0)})\right)^2},
\end{equation}
is computed and averaged over a number of simulations.  Here ${\bf
r}_i$ is the position of the $i^{th}$ particle after an appropriate
amount of simulation time, ${\bf r}_i^{(0)}$ is its initial position,
and $N$ is the number of particles.  The quantity $L$ is thus
minimized in parameter space by simulated annealing.  The program
outputs the set of parameter values that corresponds to the minimum
value of $L$, presumeably giving the potential that best suppresses
liquid nucleation and/or a possible structural phase transition.  This
is the idea of the algorithm: we postulate that if the potential is
modified in such a way that deformations of the target structure are
suppressed near (but below) the phase coexistence region, the
structure will self-assemble from a random configuration in a MC
simulation.  In the case that the melting temperature rises
significantly over the course of the optimization, the initial
temperature can be increased such that the system again approaches
coexistence.  A more detailed description of the optimization
procedure will be given in \cite{RST-pre}.  

We now apply our methodology to obtain an optimized circularly
symmetric interparticle potential, $V(r)$, that will spontaneously
favor the self-assembly of randomly placed particles in two dimensions
into the honeycomb lattice upon simulated annealing from high to zero
temperature. A claim was made that the honeycomb lattice was favored
by a hard-core plus linear-ramp potential for certain parameter values
\cite{jagla}, but we have now shown conclusively that this is not
possible because the phonon spectra of such $V(r)$ have imaginary
frequencies.  Thus, it would be a significant accomplishment for our
procedure if it assembled the honeycomb lattice.

For lattice self-assembly, we make another restriction on the
initial function: that the target should be energetically favored
among the four principal 2D lattices, (the triangular, square,
honeycomb and Kagom\'e) over a significant range of area per particle 
$\alpha$.  This is the second of our two
necessary conditions for lattice self assembly, the first being that
all phonon frequencies are real.

Both the triangular and honeycomb lattices have their first near
neighbors (in relative distances) at $1$, $\sqrt 3$ and $2$.  The
coordination numbers for those neighbors are 
6, 6, 6 and 3, 6, 3 for the triangular
and honeycomb lattices, respectively.  Thus, we choose a potential $V(r)$
such that $V(1)$ is not negative, or else there will be a tendency to
accumulate as many neighbors as a lattice will permit, and hence it
will fall into the triangular.  Yet if there is no local
minimum at the nearest neighbor distance, the mechanical stability
(phonons) will almost certainly be removed (as we have found after
trial and error).  This motivates the choice of the
family of functions 
\begin{equation}
V(r; A, \lambda) = \frac{5}{r^{12}} - \frac{6}{r^{10}} +  A\lambda^2e^{-\lambda r}.
\end{equation}

\begin{figure}
\includegraphics[scale=0.27,clip,viewport=0pt 0pt 720pt 500pt]{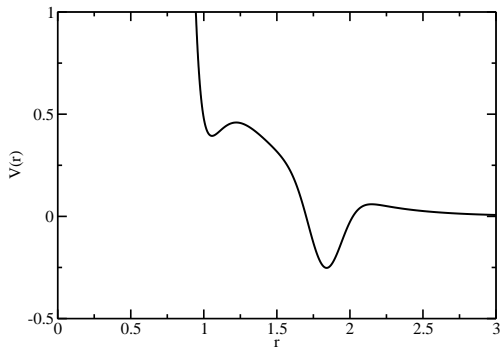}
\caption{\label{fig:hon_pot} $V(r)$, potential that favors honeycomb
lattice self-assembly.}
\end{figure}
\begin{figure}
\includegraphics[scale=0.27,clip,viewport=0pt 0pt 720pt 500pt ]{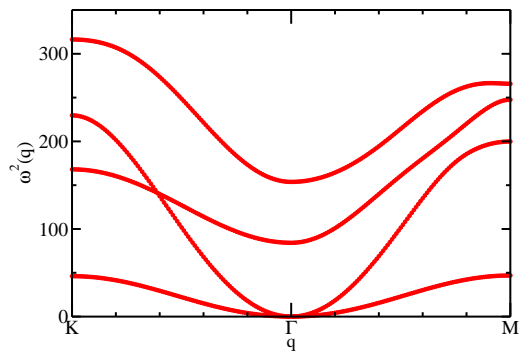}
\caption{\label{fig:hon_pho} Phonon spectrum (frequency squared) for
honeycomb with potential $V(r)$ at $\alpha = 1.45$. }
\vspace{0.2in}
\end{figure}

\begin{figure}
\includegraphics[scale=0.36,clip,viewport=0pt 0pt 450pt
450pt]{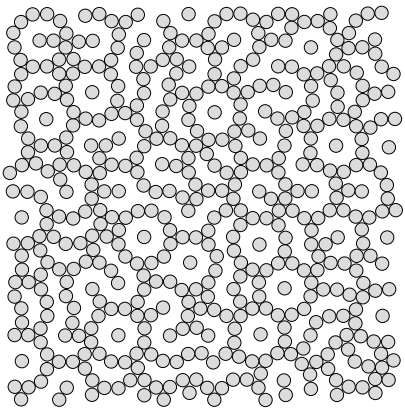}
\caption{\label{fig:hon_mc_bad} 500-particle annealed MC results, for
potential with parameters displaced from initial guess. $\alpha =
1.45$.  }
\end{figure}

\begin{figure}
\includegraphics[scale=0.36,clip,viewport=0pt 0pt 450pt 450pt]{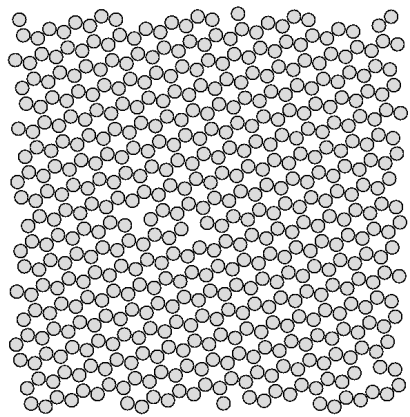}
\caption{\label{fig:hon_mc} 500-particle MC results, annealed from $k_BT = 0.22$ to $k_BT = 0$ at $\alpha = 1.45$. }
\end{figure}

This is not the function that is finally given to the
optimization program; here, we are still approximating the potential form.
This is just a LJ-type interaction (12/10 rather than 12/6) added to a
soft repulsive exponential tail that is normalized to $2\pi A$ in 2D
regardless of the value of $\lambda$.  We choose the (12/10) LJ because
it is more sharply distinguished from the exponential tail than is the
(12/6).  It also has its minimum at unity and has unit depth.  For
certain choices of $A$ and $\lambda$, this potential has one local
minimum but with a positive value.  To find the best values for $A$
and $\lambda$ to stabilize the honeycomb, we choose an $\alpha$ at
which the honeycomb nearest neighbor is at unit distance and we
maximize the interstitial hopping energy (both local and distant) of
just the exponential term over $\lambda$ at a given $A$.  We can then
set a lower bound on the value of A by constraining the potential to
have a positive hopping energy (hopping into honeycomb interstitial
sites is unfavorable).  A qualitative upper bound is set on $A$ just
by the need for mechanical stability: for large enough values of $A$,
this potential becomes purely repulsive and the honeycomb is
mechanically unstable with it.  So $A$ is chosen to be its lower bound.
This procedure gave the values $A = 3.0$ and $\lambda = 2.677$.  The
resulting potential gave favorable lattice sums but still imaginary
phonon frequencies, so in order to `brace' the lattice, an extra
attractive Gaussian was added, and so we use the paramterization
\begin{equation}
V(r; a_0, a_1, a_2, a_3) = \frac{5}{r^{12}} - \frac{a_0}{r^{10}} + a_1e^{-a_2r} - 0.4e^{-40(r-a_3)^2}.
\end{equation}

For parameter values $a_0=6.0$, $a_1=21.5$, $a_2=2.677$, and
$a_3=1.829$, the guess potential meets our two necessary conditions.
A 500-particle annealed MC simulation using this potential produced a
lattice reminiscent of the honeycomb, but with a significant number of
defects in the ground state.

In order to demonstrate the effectiveness of the optimization program,
we somewhat arbitrarily displaced the parameters from the initial
guess function, setting them to be $a_0=6.5,\ a_1=18.5, \ a_2=2.45, \
a_3=1.83$.  The resulting 500-particle annealed configuration 
is shown in Fig. \ref{fig:hon_mc_bad}.  We then started
the optimization program with these values for the parameters, the
output of which was the following potential:
\begin{equation}
V_{f}(r) = \frac{5}{r^{12}} - \frac{5.89}{r^{10}} + 17.9e^{-2.49r} - 0.4e^{-40(r-1.823)^2}.\label{eq6}
\end{equation}  

This function is plotted in Fig. \ref{fig:hon_pot}.  The phonon
spectrum is given in Fig. \ref{fig:hon_pho}.  The lattice sums (not
shown there) demonstrate that while the honeycomb is stable over a
wide range of $\alpha$, the triangular eventually dips below it.  The
global minimum for the triangular occurs at the area at which the
first nearest neighbor lies at the bottom of the Gaussian.  As long as
there is an attractive gaussian in $V(r)$, there is no avoiding this
effect.  However, it has little relevance for the lattice
self-assembly, since it is at relatively high $\alpha$.

Monte Carlo simulated annealing was carried out on a 500-particle
system (at $\alpha = 1.45$) interacting via the potential shown in
Fig. \ref{fig:hon_pot} (equation \ref{eq6}).  The resulting configuration is depicted
Fig. \ref{fig:hon_mc}.  Except for a few defects, the honeycomb has
indeed self-assembled.  The defects actually seem to be ``missing
particles," rather than dislocations or pockets of disordering.  
One might expect that increasing the number of particles to fill the defects
(or adjusting $\alpha$ accordingly) would eliminate the defects, but this
is not the case.  The defects are likely due to the slow
dynamics of the MC, i.e., they were ``frozen in" during annealing.
According to the lattice sums, the perfect honeycomb lattice is lower
in energy than the one produced in the MC simulation with defects.  

We have found that as long as the salient features of the honeycomb
potential are kept (two local minima at distance ratio $\sqrt{3}$,
the first being positive and the second negative), self-assembly is
unaffected by perturbations in the potential, i.e. the potential
is robust.  This is essential if this system is to be tested
experimentally.

In summary, using an inverse statistical-mechanical approach, 
we have found an optimized isotropic pair potential that results in the
self-assembly of the targeted honeycomb lattice.  In many
nanoscopic systems, experimentalists have increasingly greater control
over inter-component interactions, and hence, optimal design of
nanostructures by self-assembly ultimately is always an inverse problem.  
Our results give some hope to the possibility of
self-assembly of the more challenging diamond lattice
with isotropic pair potentials. As we indicated earlier, this
would have important implications for photonics devices.
We are currently searching for a parameterization that upon
optimization of an isotropic pair potential will yield the
diamond crystal as its ground state. Note that our
methodology also offers the opportunity to examine defects in
these ideal structures and to minimize their occurrence by maximizing
their energy costs.

There are many fascinating research avenues that we can explore using
our inverse approach. Our results beg the question of whether more
exotic structures can be assembled using only isotropic pair
potentials.  For example, one might try to assemble a Buckyball in a
NVT annealing simulation with 60 particles with a potential that has
sharp minima at the first several neighbor positions.  Although we do
not know whether this is possible, we have already found a potential
that produces small clusters of particles (e.g., simplices), as well
as one that produces long chains, or ``colloidal nanowires."  Clearly,
there must be limits on the types of structures that can be assembled
using only isotropic potentials. We know, for example, that chiral
structures with a target chirality cannot be formed, but beyond this
specific case we know very little about the limitations of isotropic
pair potentials: a fundamentally important problem.

The optimization algorithm proposed in this Letter is only one
approach to the inverse problem, and we expect that others will be
needed to search for interactions  (isotropic or not, additive or not)
that stabilize non-periodic systems.  Apart from any particular
algorithm, however, a central point of this Letter is to propose the
use of powerful inverse statistical-mechanical techniques to
exquisitely control self-assembly from the nanoscopic to microscopic
scales.

This work was supported by the Office of Basic Energy Sciences, DOE, 
under Grant No. DE-FG02-04ER46108.

\end{document}